\documentclass[a4paper,twoside]{article}
\newcommand{\affil}[1]{$^{\rm #1}$}
\usepackage[authoryear]{natbib}
\bibpunct{(}{)}{;}{a}{}{,}
\usepackage{graphicx}
\usepackage[colorlinks,linkcolor=blue,citecolor=black,hyperindex,CJKbookmarks,dvipdfm]{hyperref}
\date{} 
\title{\large\bf\flushleft Study of The Abundance Patterns in The Metal-Poor Stellar Stream}
\author{\parbox{\textwidth}{\flushleft
\vspace{-0.5cm}
{\it Hongjie Li\affil{A,B,C}, Shuai Liang\affil{A,B}, Wenyuan Cui\affil{A,B} and Bo Zhang\affil{A,B,D}}\\
\vspace{0.4cm}
{\small \affil{A}\,Department of Physics, Hebei Normal University,
113 Yuhua Dong Road, Shijiazhuang 050016, China} \\
{\small \affil{B}\,Hebei Advanced Thin Films Laboratory, Shijiazhuang 050016, China}\\
{\small \affil{C}\,School of Sciences, Hebei University of Science
and Technology, Shijiazhuang 050018, China}\\
{\small \affil{D}\,Corresponding author. Email:
zhangbo@mail.hebtu.edu.cn}}}
\begin{document}
\twocolumn[
\begin{changemargin}{.8cm}{.5cm}
\begin{minipage}{.9\textwidth}
\vspace{-1cm}
\maketitle
%
%
\small{\bf Abstract:} The chemical abundances of the metal-poor
stars in the stellar stream provide important information for
setting constraints on models of neutron-capture processes. The
study of these stars could give us a better understanding of
r-process nucleosynthesis and chemical composition of the early
Galaxy. Using the updated main r-process and weak r-process
patterns, we fit abundances in the stellar stream stars. The weak
r-process component coefficients are almost constant for the
sample stars, including r-rich stars, which means that both weak
r-process and Fe are produced as primary elements from SNeII and
their yields have nearly a constant mass fraction. The difference
between the stream stars and r-rich stars is obvious. For the
stream stars, that the increase trend in the main r-process
component coefficients as metallicity increases means the gradual
increase in the production of main r-process elements relative to
iron. This behavior implies that the masses of progenitors for the
main r-process are smaller than those of the weak r-process.
Furthermore, we find metal-poor stream star HD 237846 is a weak
r-process star.


\medskip
\medskip
\end{minipage}
\end{changemargin}
]
\small

\section{Introduction}

The formation and evolution of Galaxy have not been fully
understood, although many studies have explored the problems for
several decade years. Early studies thought that our Galaxy formed
through isolated collapse \citep{egg62}. The idea that the halo of
the Galaxy may have formed partly by accreting some small fragments
was proposed by \cite{sea78}. The stars related to these fragments
could be recognizable as kinematic substructures and chemical
composition among the Galactic halo stars. They should be
concentrated into a number of coherent ``streams" in velocity space.
\cite{hel99} discovered a stellar steam in the solar neighborhood by
detecting the distribution of angular momentum components. They
estimated that about ten per cent metal-poor halo stars came from
the stellar stream whose progenitor system was disrupted during the
formation of the Galaxy. Because the abundance ratios of
[$\alpha$/Fe] in the stream are similar to those of other halo
stars, \cite{kep07} suggested that the abundances of these stream
stars are mostly enriched by Type II supernovae. The stellar stream
provides an excellent test bed for understanding the formation of
our Galaxy. In this aspect, a detail study of elemental abundances,
including the abundances of neutron-capture elements, in the stream
is important.

The elements heavier than the iron-peak elements can be made through
two neutron-capture processes: the r-process and the s-process
\citep{bur57}. The s-process is further divided into two components,
i.e., the main s-component and the weak s-component. The sites of
main s-component are thought to be in low- to intermediate-mass
stars in the asymptotic giant branch (AGB) \citep{bus99}. The most
likely sites for the weak s-component are in core helium burning and
shell carbon burning in massive stars. The r-process sites are
associated with Type II supernovae (SNe) from massive stars, but it
is not fully confirmed \citep{cow91,sne08}. Observations of the very
metal-poor halo stars CS 22892-052 \citep{sne03,cow05} and CS
31082-001 \citep{hil02,hon04} have revealed that their heavier
neutron-capture elements (Z $\geq 56$) are in remarkable agreement
with the scaled solar-system r-process pattern produced by the main
r-process \citep{cow99,tru02,wan06}, but their lighter
neutron-capture elements ($37\leq Z\leq47$) seemed to be deficient
\citep{sne00,hil02}. This means that the r-process abundance pattern
in solar-system material can not be explained by a single process.
Based on the calculation of Galactic chemical evolution,
\cite{tra04} concluded that another component from massive stars is
needed to explain the solar abundance of Sr, Y and Zr. The
corresponding process has the primary nature and is called as
``lighter element primary process" (``LEPP", see \cite{tra04} and
\cite{ser09}) or ``weak r-process component''
\citep{ish05,cow05,cow06,mon07,izu09}. \cite{zha10} analyzed the
abundances of 14 metal-poor stars with the metallicities
[Fe/H]$<-2.0$ and found that the abundance patterns of both
neutron-capture elements and light elements could be best explained
by having a star form in a molecular cloud polluted by weak r- and
main r-process material.

A quantitative understanding of the origins of r-process elements in
the metal-poor stars is still difficult. In this aspect, the stream
stars are optimum sites for the study of r-process because most
stream stars have not been polluted by the s-process material.
Recently, \cite{roe10} observed the abundances of 12 metal-poor
stream stars. Their angular momentum components clumped together far
more than the random distribution of halo stars, which was first
reported by \cite{hel99}. All of 12 members have angular momentum
components with 500$<L_{z}<1500$ kpc km $s^{-1}$ and
$L_{\bot}\sim2000$ kpc km $s^{-1}$, where $L_{z}$ is the angular
momentum component of rotation for a given star's orbit and
$L_{\bot}$ is the combination of the other two angular momentum
components ($L_{\bot}=\sqrt{L_{x}^{2}+L_{y}^{2}}$). Based on
observations of metal-poor stream stars, \cite{roe10} concluded that
the observed abundance of the heavy elements in most stream stars
cannot be well fitted by the abundances from a single r-process,
especially for the lighter neutron-capture elements Sr, Y and Zr. In
order to better understand the origins of neutron-capture elements
in the stellar stream, one must consider the contribution of weak
r-process. In order to make progress in our understanding of the
neutron-capture processes, we study the element abundance patterns
of the metal-poor stream stars, in which light elements, iron group
elements, lighter and heavier neutron-capture elements are observed.

In this paper, we calculate the relative contributions from the
individual neutron-capture process to the elemental abundances of 9
stream stars in which elemental abundances of Eu have been observed.
In Section 2, we obtain the two r-process components and extend the
two abundance patterns to light elements. Using the updated weak
r-component and main r-component, we calculate the relative
contributions from the individual neutron-capture process to the
elemental abundances in the stream stars. The calculated results are
presented in Section 3. Our conclusions are given in Section 4.

\section{The Weak r-process and Main r-process Components}

Observational abundances in metal-poor stars show that sites
producing the heavy r-nuclei do not produce iron group elements or
the other light elements heavier than N \citep{qia07}. In this
fact, the ``main r-process stars" CS 22892-052 and CS 31082-001
could have abundances that mainly reflect results of the main
r-process nucleosynthesis of a few SNe. On the other hand, the
``weak r-process stars", i.e. HD 122563 and HD 88609, play an
essential role in constraining the weak r-process, since they have
the smallest contribution from the main r-process. Adopting the
abundances of CS 22892-052 as main r-process pattern, \cite{zha10}
found that, although the pattern of heavier neutron-capture
elements are very similar to those main r-process pattern, the
light-element abundance patterns of r-rich stars are close to
those of weak r-process stars. In this study, based on the detail
abundance analysis of main r-process stars and weak r-process
stars, we find that the abundances of light and iron group
elements in main r-process stars do not come from main r-process
events. We subtract the average abundances of main r-process stars
from the average abundance of weak r-process stars, normalized to
Eu. And then subtract the obtained result from the average
abundance of main r-process stars, normalized to Fe. Repeat the
processes above, we can obtain the pure weak r- and main
r-components, which had been scaled to the solar r-process
abundances given by \cite{arl99} and the values of Sr-Nb updated
from \cite{tra04}.

\section{Results and Discussions}

\subsection{Fitted Results}

\begin{figure*}[h]
\begin{center}
\includegraphics[scale=1.5, angle=0]{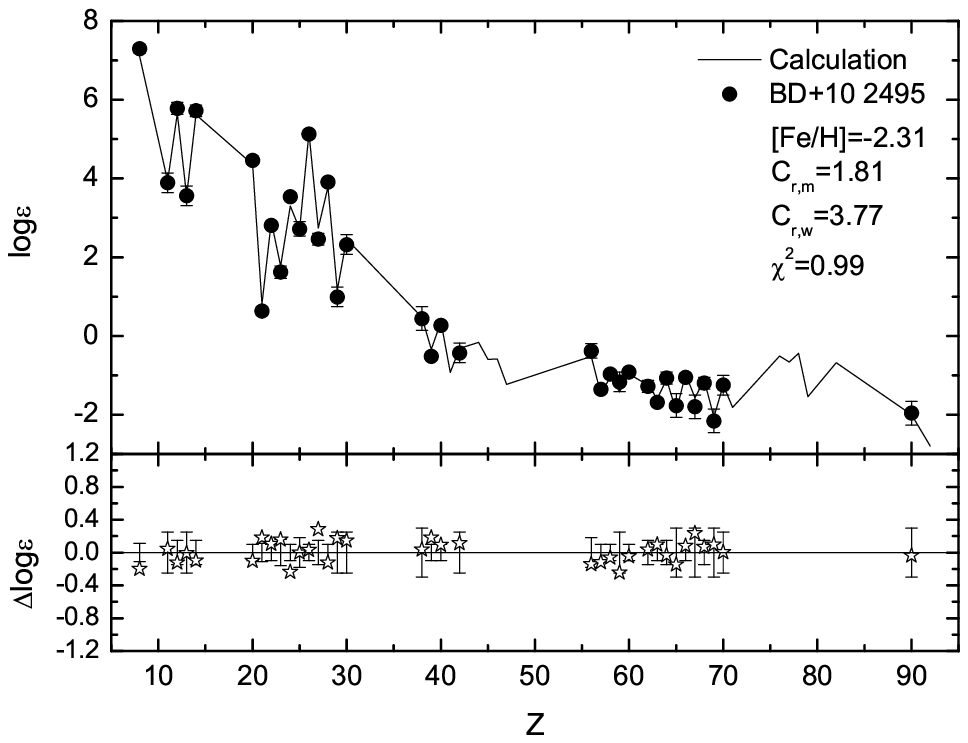}
\caption{An example for the best-fit results of stream stars. In the
top panel, the solid line presents our calculated result. The filled
circles are observed elemental abundances. In the bottom panel, the
open stars are individual relative offsets
($\Delta\log\varepsilon(X)\equiv \log\varepsilon(X)_{cal}
-\log\varepsilon(X)_{obs}$). The error bars are the observational
errors.}\label{fig1}
\end{center}
\end{figure*}

The observed abundances of nine stream stars are taken from
\cite{roe10}. We will explore the origin of the neutron-capture
elements in the stars, by comparing the observed abundances with
predicted results. The $i^{th}$ element abundance in a star can be
calculated as
\begin{equation}
N_{i}(Z)=(C_{r,m}N_{i,r,m}+C_{r,w}N_{i,r,w})\times10^{[Fe/H]}
\end{equation}
where $N_{i,r,m}$, $N_{i,r,w}$ are the scaled abundances of the
$i^{th}$ element produced by the main r-process and weak r-process,
respectively; $C_{r,m}$ and $C_{r,w}$ are the component coefficients
corresponding to relative contributions from the main r-process and
weak r-process to the elemental abundances, respectively. Based on
equation (1), using the observed data of stream stars \citep{roe10},
we can obtain the best-fit $C_{r,m}$ and $C_{r,w}$ by looking for
the minimum $\chi^2$. The component coefficients and $\chi^2$ are
listed in Table \ref{table1}. Fig. \ref{fig1} shows an example of
our calculated best-fit results for the stream stars. The observed
elemental abundances marked by filled circles are also shown to
facilitate comparison. For most stars, there is good agreement
between the predictions and the data for most elements starting with
O to Pb in consideration of the observational errors. \cite{roe10}
have suggested that the abundances of the stream stars do not
contain the contribution of s-process, except for s-rich star CS
29513-032. Furthermore, because of the effect of the secondary-like
nature of the major neutron source in massive stars, no weak
s-contribution is expected in halo stars \citep{tra04}. As a test,
adding the main s-process component and weak s-process component to
equation (1), we use four components to fit the abundances of
metal-poor stars BD+29 2356 ([Fe/H]=-1.59) and BD+30 2611
([Fe/H]=-1.50). The main s-process pattern is taken from the result
with [Fe/H]=-2.0 calculated by \cite{bus01} and weak s-process
pattern is taken from the result presented by \cite{rai93}. The
components coefficients deduced for BD+29 2356 are $C_{r,m}$=3.85,
$C_{r,w}$=4.14, $C_{s,m}$=0 and $C_{s,w}$=0; for BD+30 2611,
$C_{r,m}$=4.49, $C_{r,w}$=2.82, $C_{s,m}$=0 and $C_{s,w}$=0.11. We
can see that the component coefficients of main s-process and weak
s-process are smaller by one order of magnitude than those of main
r- and weak r-process. Our results are consistent with suggestion
presented by \cite{roe10} and \cite{tra04}.

\begin{figure*}[h]
\begin{center}
\includegraphics[scale=1.1, angle=0]{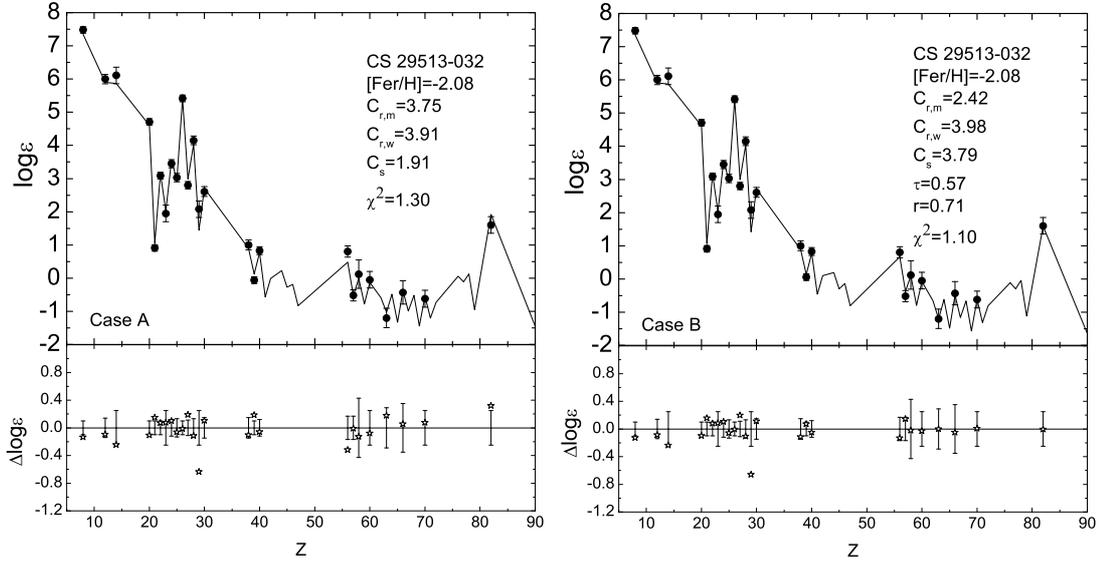}
\caption{The best-fit results for CS 29513-032. The left panel shows
our result by adopting the result of AGB model with [Fe/H]=-3 given
by \cite{bus01}. The right panel shows our result by adopting the
parametric approach. The symbols are the same as in Fig.
\ref{fig1}.}\label{fig2}
\end{center}
\end{figure*}

In the stream stars, CS 29513-032 shows clear enrichment by the
s-process. \cite{roe10} have speculated that the s-process material
observed in CS 29513-032 is accreted from another low-metallicity
AGB star. If it is the case, the s-process abundance in the envelope
of the stars could be expected to be lower than the abundance
produced by the s-process in AGB star because the material is mixed
with the envelopes of the primary (former AGB star) and secondary
stars. \cite{roe10} found that the observed abundances could not be
fitted by the scaled solar system r-process pattern nor by the
s-process pattern. We explore the origin of the neutron-capture
elements in the s-rich stars by comparing the observed abundances
with predicted s- and r-process contribution. After considering
s-process contribution, the $i^{th}$ element abundance in the
envelope of the star can be calculated as follows:
\begin{equation}
N_{i}(Z)=(C_{s}N_{i,s}+C_{r,m}N_{i,r,m}+C_{r,w}N_{i,r,w})\times10^{[Fe/H]}
\end{equation}
where $C_{s}$ is the component coefficients that correspond to
s-process and $N_{i,s}$ is the abundance produced by the s-process
in AGB star, which has been normalized to the s-process abundance of
Ba in the solar system \citep{arl99}. The adopted abundance pattern
$N_{i,s}$ in equation (2) is taken from the main s-process abundance
with [Fe/H]=-3.0 for 1.5$M_{\odot}$ given by \cite{bus01} (see their
Fig. 2). The fitted result (Case A) is presented in the left panel
of Fig. \ref{fig2}. We can see that, for most observed
neutron-capture elements, their abundances can be fitted within the
observational uncertainty. This means that the s-process elements in
CS 29513-032 come from the dredged-up material in the former
low-mass AGB star. The r-process coefficients and $\chi^{2}$ are
$C_{r,m}$=3.75, $C_{r,w}$=3.91 and $\chi^{2}$=1.3. We can see that
its r-process coefficients are close to those of other stream stars
with similar metallicity. It is interesting to adopt the parametric
model presented by \cite{aok01} and developed by \cite{zha06} for
s-rich stars to study the physical parameters which could reproduce
the observed abundances. There are five parameters in the parametric
model. They are the neutron exposure $\tau$ , the overlap factor
$r$, the component coefficient of the s-process $C_{s}$ and the
component coefficients of the two r-processes $C_{r,w}$, $C_{r,m}$.
Using the parametric approach, we can carry out s-process
nucleosynthesis calculation to fit the abundance profile observed in
the s-rich stars by look for the minimum $\chi^{2}$ in the
five-parameter space formed by $\tau$ , $r$, $C_{s}$, $C_{r,w}$ and
$C_{r,m}$. The right panel (Case B) in Fig. \ref{fig2} shows our
best-fit result. Because the parametric model contains more
parameters, the $\chi^{2}$ is smaller than that of Case A.

\begin{figure*}[h]
\begin{center}
\includegraphics[scale=1, angle=0]{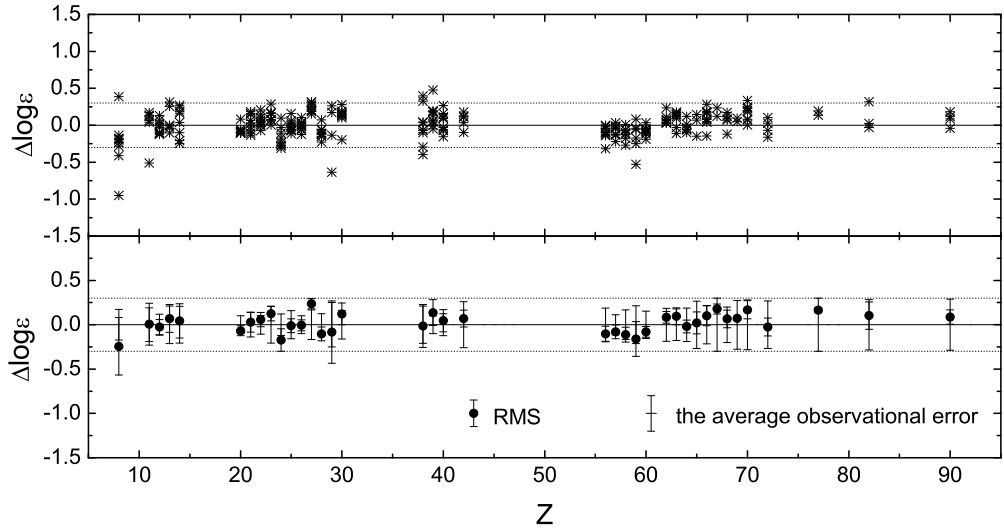}
\caption{Top panel: individual relative offsets
($\Delta\log\varepsilon(X)$, $\Delta\log\varepsilon(X)\equiv
\log\varepsilon(X)_{cal} -\log\varepsilon(X)_{obs}$) for the sample
stars with respect to the predictions from the abundance model
(stars). Typical observational uncertainties in $\log\varepsilon$
are $\sim0.2-0.3$ dex (dotted lines). Bottom panel: The
root-mean-square offsets of these elements in $\log\varepsilon$.
Filled circles are average stellar abundance offsets.}\label{fig3}
\end{center}
\end{figure*}

In the top panel of Fig. \ref{fig3}, we show individual relative
offsets ($\Delta\log\varepsilon$) for the sample stars with respect
to the predictions. Typical observational uncertainties in
$\log\varepsilon$ are $\sim0.2-0.3$ dex (dotted lines). The
root-mean-square offsets (RMS) of these elements in
$\log\varepsilon$ shown in the bottom panel are mostly smaller than
0.30 dex for the comparison with the calculated results. These
values are consistent with zero, given by the combined uncertainties
in stellar abundances and predicted abundances, which confirm the
validity of the approach adopted in this work. From Fig. \ref{fig1}
to Fig. \ref{fig3} we can see that the agreement are obtained for
both light elements, iron group elements and neutron-capture
elements.

\subsection{The Two r-process Component Coefficients}

\begin{figure*}[h]
\begin{center}
\includegraphics[scale=1.5, angle=0]{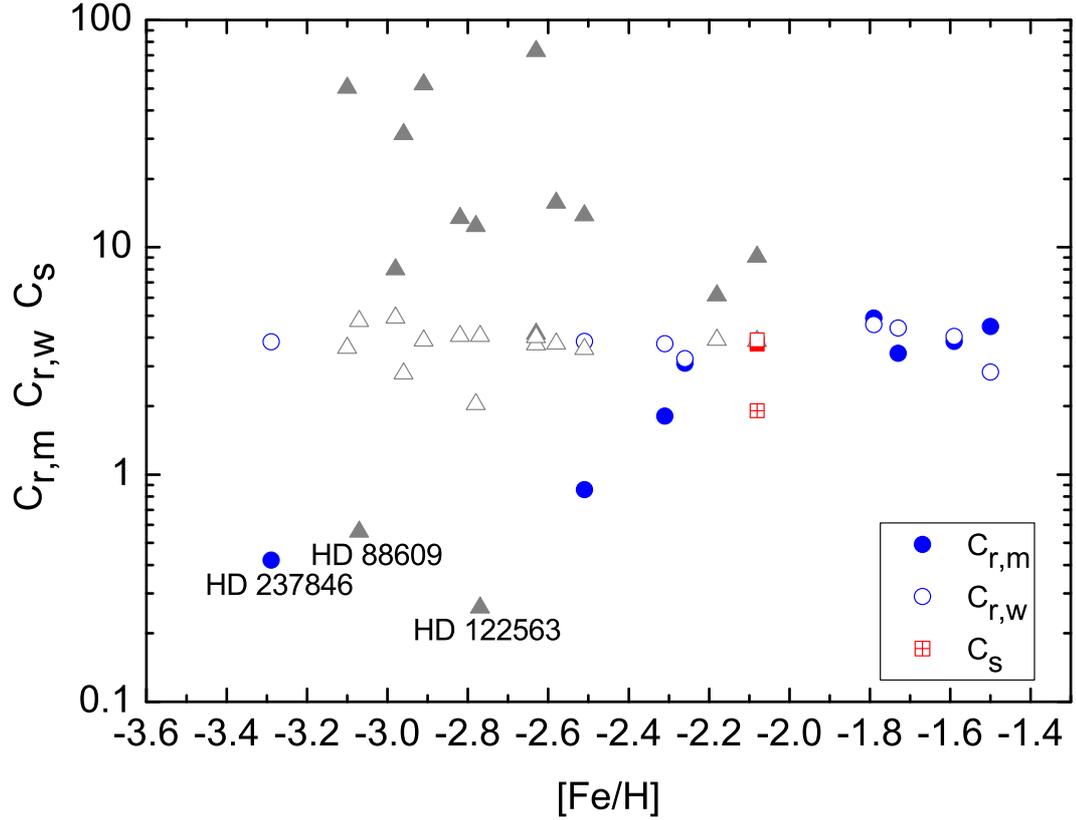}
\caption{The component coefficients as a function of metallicity.
Symbols: filled circles and open circles are the main r-process
coefficients and the weak r-process coefficients of 8 stream
stars, respectively. Filled squares, open squares and open squares
with across are the component coefficients responsible for the
maim r-process, weak r-process, and the main s-process in s-rich
star CS 29513-032, respectively. Filled triangles and open
triangles are the main r-component coefficients and the weak
r-component coefficients of 14 metal-poor halo stars studied in
\citep{zha10}, respectively. Their abundances are observed by
\cite{wes00,cow02,hil02,sne03,hon04,hon06,hon07,bar05,iva06,chr08,hay09,mas10,roe10b}.}\label{fig4}
\end{center}
\end{figure*}

The component coefficients as a function of metallicity, illustrated
in Fig. \ref{fig4}, contain some important information about the
polluted history of the stellar stream. The trend of coefficients
$C_{r,w}$ is almost constant for the stream stars, which is clearly
different from that of $C_{r,m}$. The calculated results that
$C_{r,w}$ is nearly constant suggest that the elements produced by
weak r-process have increased along with Fe over the polluted
history of the stellar stream. This means that both weak r-process
elements and Fe are produced as primary elements from SNe II and
their yields have nearly a constant mass fraction. \cite{roe10} have
found that most of $\alpha$-elements and iron group elements have
small or moderate dispersions, and show no evolution with
metallicity. This also means that both light elements and iron group
elements are produced as primary elements from SNe II in which weak
r-process occur, and their yields have nearly a constant mass
fraction. On the other hand, they found that most of neutron-capture
elements follow similarly increasing trends of [element/Fe] ratios
with increasing [Fe/H]. Obviously, the increase trend in $C_{r,m}$
as [Fe/H] increases is a consequence of the gradual increase in the
production of main r-process elements relative to iron. The rise
means that the main r-process contribution may begin slightly later
than the contribution from weak r-process producing in conjunction
with the Fe. This behavior supports the idea that the masses of
progenitors for the main r-process are smaller than those of the
weak r-process (or LEPP)
\citep{tra99,tra01,tra04,wan06,qia07,izu09}. The main r-process
coefficient $C_{r,m}$ is close to the weak r-process coefficient
$C_{r,w}$ as the metallicity approaches [Fe/H]$\sim-2.2$.

For ease of comparison, the recalculated coefficients of 12 r-rich
stars and two weak r-process stars (HD122563 and HD 88609) studied
in \cite{zha10} are marked by triangles in Fig. \ref{fig4}. We
find that their weak r-process component coefficients are close to
those of the stream stars. However, for main r-process components,
the difference between the stream stars with [Fe/H]$<-2.2$ and
r-rich stars is obvious. As seen from Fig. \ref{fig4}, the sample
stars are separated into two branches clearly by $C_{r,m}$, i.e.,
r-rich stars having coefficient $C_{r,m}$ larger than 3.0 and
``$low-C_{r,m}$'' stars with coefficient $C_{r,m}$ less than 3.0.
All metal-poor stream stars ([Fe/H]$<-2.2$) and two weak r-process
stars (HD 122563 and HD 88609) belong to ``$low-C_{r,m}$'' stars.

\subsection{Discussion of The Abundance Decomposition}

\begin{figure*}[h]
\begin{center}
\includegraphics[scale=1.5, angle=0]{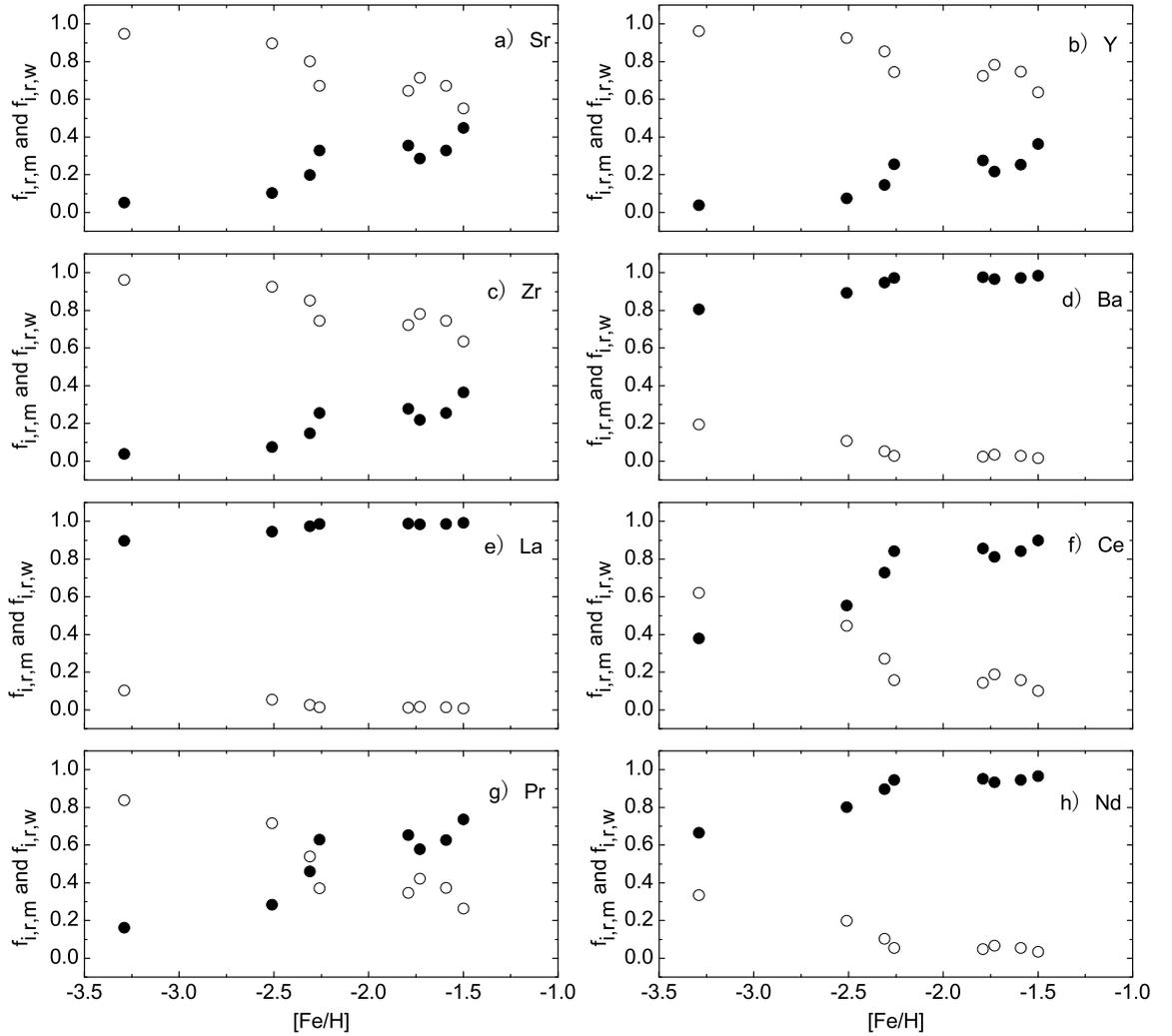}
\caption{The component fractions of 8 stream stars as function of
[Fe/H]. Symbols: filled circles and open circles are the fractions
of main r-component and the weak r-component,
respectively.}\label{fig5}
\end{center}
\end{figure*}

As the abundances of the stream stars can be used as a probe of
the origin of the neutron-capture elements in their progenitor
system, it is important to determine the relative contributions
from the individual neutron-capture processes to the abundances of
heavy elements in the stars. It was possible to isolate the
contributions corresponding to the weak r- and main r-processes
using our approach. Taking the values of $C_{r,m}$ and $C_{r,w}$
into equation (1), the decomposed results of sample stars can be
derived. In Fig. \ref{fig5} (a)-(h), the component fractions
$f_{i,r,m}$
\begin{math}
(f_{i,r,m}=\frac{C_{r,m}N_{i,r,m}}{C_{r,m}N_{i,r,m}+C_{r,w}N_{i,r,w}})
\end{math}
and $f_{i,r,w}$
\begin{math}
(f_{i,r,w}=\frac{C_{r,w}N_{i,r,w}}{C_{r,m}N_{i,r,m}+C_{r,w}N_{i,r,w}})
\end{math}
as function of [Fe/H] are given by the filed circles (main
r-process) and open circles (weak r-process), respectively. The
abundances of Sr, Y and Zr in the stream stars are a mixture of two
r-process components. We can see that, for the sample star with the
lowest metallicity, i.e. [Fe/H]=-3.29, the weak r-processes are
dominantly responsible for abundances of the lighter neutron-capture
elements Sr, Y and Zr. Although the contributions from weak
r-processes are larger than those of main r-process to the lighter
neutron-capture elements of the stream stars, the contributions from
the main r-process to lighter neutron-capture elements, such as Sr,
Y and Zr, increase with increasing metallicity.

\begin{figure*}[h]
\begin{center}
\includegraphics[scale=1.5, angle=0]{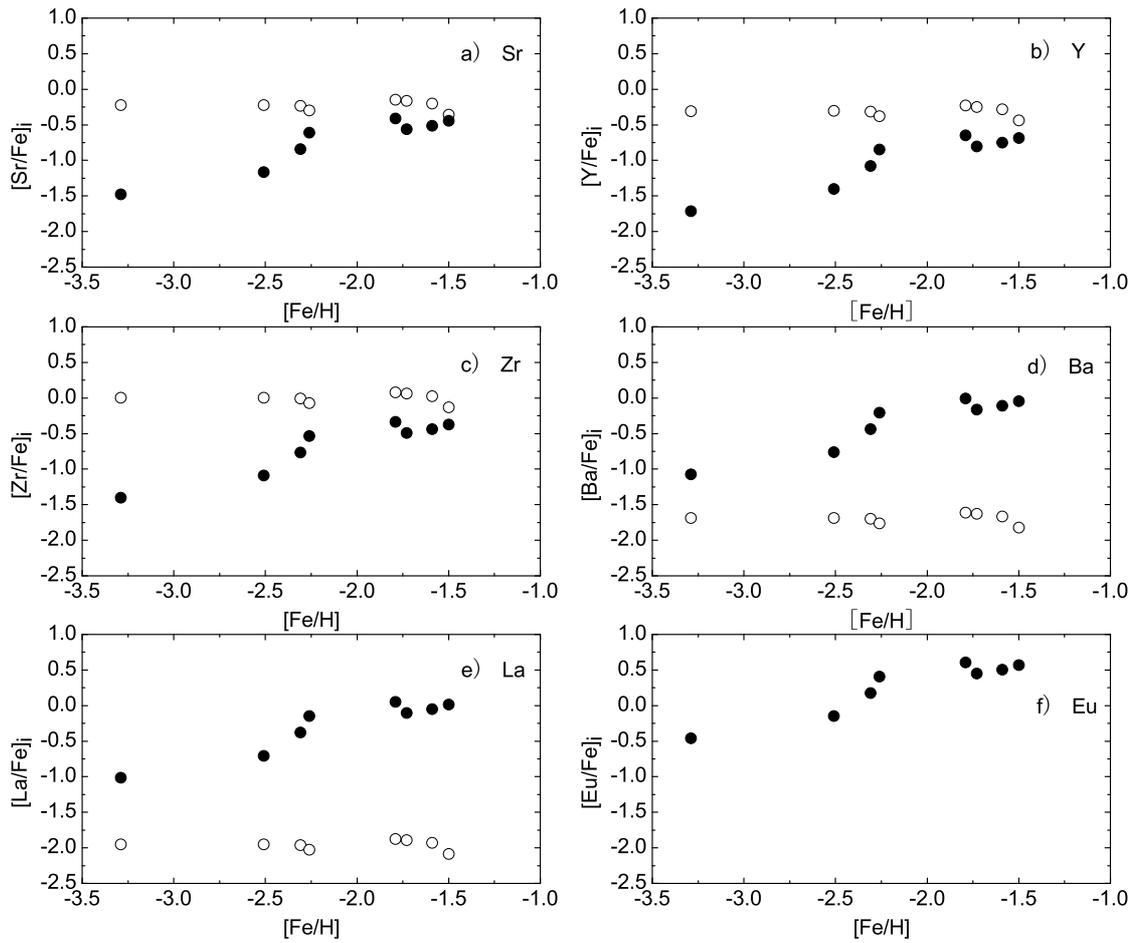}
\caption{The logarithmic component ratios vary with metallicity for
8 stream stars. Symbols: filled circles and open circles are the
main r-component ratios and the weak r-component ratios,
respectively.}\label{fig6}
\end{center}
\end{figure*}

In order to study the origins of neutron-capture elements, the
variation of the logarithmic ratio [element/Fe] with metallicity is
particularly useful. The component ratios of the individual
neutron-capture process, i.e., [element/Fe]$_{i}$ (i =r,w; r,m) are
derived and shown in Fig. \ref{fig6}. The filled circles and open
circles respectively represent the main r- and weak r-component
ratios. The trend of ratios [element/Fe]$_{r,w}$ is almost constant
for the stream stars, which is clearly different from that of
[element/Fe]$_{r,m}$. In particular, we focus on the ratios of Sr,
Y, Zr, which are partly produced by weak r-process nucleosynthesis.
We can see that, for the sample star, the contributions from weak
r-processes are higher than those of main r-process for abundances
of the lighter neutron-capture elements Sr, Y and Zr. The calculated
results that the [Sr,Y,Zr/Fe]$_{r,w}$ is nearly constant suggest
that the elements produced by weak r-process have increased along
with Fe. This also means that both weak r-process and Fe are
produced as primary elements from SNe II and their yields have
nearly a constant mass fraction. From Fig. \ref{fig6}(a) we find
that the ratios [Sr/Fe]$_{r,w}$ are nearly -0.3. For the other
lighter neutron-capture elements Y and Zr, the similar results can
be obtained. To investigate the contribution of the main r-process
in [element/Fe] for heavier neutron-capture elements Ba, La and Eu,
the variations of the logarithmic ratios [element/Fe]$_{i}$ are
shown in Fig. \ref{fig6}(d)-(f). For the heavier neutron-capture
elements, the figures show that the main r-process contribution is
dominant. We find that a increase trend in the [element/Fe]$_{r,m}$
as [Fe/H] increases. This would mean a consequence of the gradual
increase in the production rate of main r-process relative to iron
due to the increased contribution of SNe II in which main r-process
events occur. Our calculated results could be used as the
constraints on the chemical characteristics of progenitor system
from which the stream stars originated.

\begin{figure*}[h]
\begin{center}
\includegraphics[scale=1.5, angle=0]{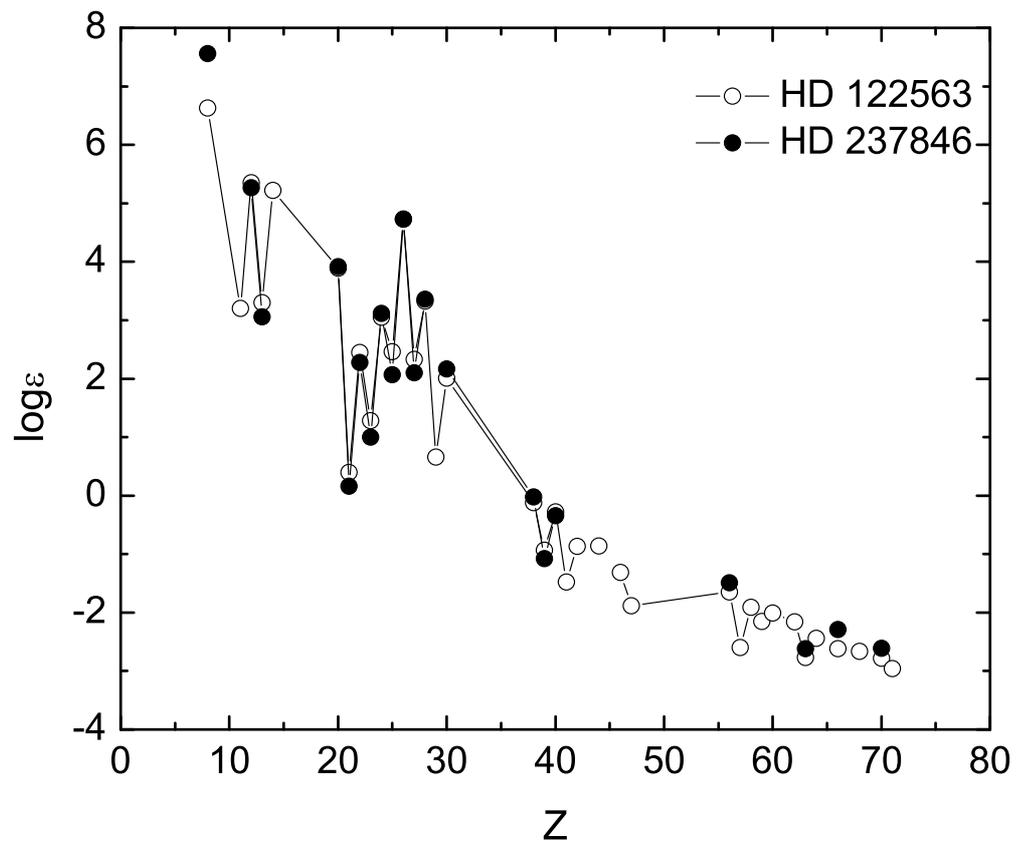}
\caption{The abundance comparisons on the logarithmic scale between
HD 237846 and HD 122563 as a function of atomic number. Symbols: the
open circles and filled circles present the observed abundances in
HD 122563 and HD 237846, respectively.}\label{fig7}
\end{center}
\end{figure*}

Up to now, the metal-poor stars HD 122563 and HD 88609 are only two
stars having extremely excesses of lighter neutron-capture elements
and are called as ``weak r-process stars". Their overall abundance
pattern could represent the yields of the weak r-process. In order
to investigate the robustness of weak r-process pattern, it is very
important to find the more weak r-process stars. The component
coefficients can be used to select special stars. If one component
coefficient is much larger than others, the abundances of this star
may be dominantly produced by the corresponding process. Based on
our results shown in Fig. \ref{fig4} and Table \ref{table1}, we can
find that the metal-poor stars, HD 237846 with $C_{r,m}=0.42$,
$C_{r,w}=3.84$ could be one weak r-process star. In order to test
this finding, Fig. \ref{fig7} shows the abundance comparisons on the
logarithmic scale between HD 237846 and HD 122563 as a function of
atomic number, which have been normalized to Fe abundance of HD
122563. Our conclusion here is that, except element O, the abundance
patterns of HD 237846 and HD 122563 are quite similar. The
metallicity of HD 237846 ([Fe/H]=-3.29) is even lower than that of
HD 88609. Our present study reveals that HD 237846 is another weak
r-process star.

\section{Conclusions}

In our Galaxy, nearly all chemical evolution and nucleosynthetic
information is in the form of elemental abundances of stars with
various metallicities. In this aspect, the stellar abundances
polluted by only a few processes, such as main r-process stars and
weak r-process stars, are very significant, because their abundances
can be directly compared with model prediction. However, such stars
are very rare. The chemical abundances of the metal-poor stars in
the stellar stream are excellent information for setting constraints
on models of r-processes. Our results can be summarized:

1. The observed abundances of the heavy elements in the stream stars
cannot be well fitted by the abundances of the main r-process
\citep{roe10}. After adding the contribution of weak r-process and
its accompaniment, the abundances of most metal-poor stream stars,
including light elements, iron group elements and neutron-capture
elements, can be fitted. For lighter neutron-capture elements Sr, Y
and Zr, the contributions from weak r-process are larger than those
of main r-process.

2. The weak r-process coefficients are almost constant for the
sample stars, including r-rich stars. This means that the elements
produced by weak r-process have increased along with Fe over the
polluted history of the stellar stream. Both weak r-process and Fe
are produced as primary elements from SNeII and their yields have
nearly a constant mass fraction.

3. The increase trend in the [element/Fe]$_{r,m}$ as [Fe/H]
increases means the gradual increase in the production rate of
main r-process relative to iron. This behavior is due to the
increased contribution of SNe II in which main r-process events
occur and would imply that the masses of progenitors for the main
r-process are smaller than those of the weak r-process.

4. The difference between the stream stars and r-rich stars is
obvious. The r-rich stars and the stream stars are separated into
two branches clearly, with a systematic different in main
r-process component $C_{r,m}$ at a given metallicity. All
metal-poor stream stars ([Fe/H]$<-2.2$) belong to
``$low-C_{r,m}$'' stars. Moreover, the weak r-process stars HD
122563 and HD 88609 also belong to ``$low-C_{r,m}$'' stars.

5. Although large number of metal-poor stars have been studied
over the past few decades, there are only two weak r-process
stars, HD 122563 ([Fe/H]=-2.77) and HD 88609 ([Fe/H]=-3.07), have
been found. Using the component coefficients, chemical specific
stars can be selected. We find that the metal-poor star HD 237846
is another weak r-process star. The metallicity of HD 237846 is
[Fe/H]=-3.29, which means that abundance pattern produced by weak
r-process is stable from star to star for a wider range of
metallicity. This result would give stronger constraint on the
models of the weak r-process in the very early Galaxy.

6. The calculated results imply that the relative contribution
from the individual neutron-capture process to the heavy element
abundances is not usually in the solar proportion, for most stream
stars. The contributions of the weak r-process are larger than
those of the main r-processes to lighter neutron-capture elements
Sr, Y and Zr. The main r-process coefficient $C_{r,m}$ close to
the weak r-process coefficient $C_{r,w}$ as the metallicity
approaches [Fe/H]$\sim-2.2$, which implies that both the weak
r-process and the main r-process component fractions reach their
solar-system values as the metallicity approaches
[Fe/H]$\sim-2.2$.

7. CS 29513-032 is an only s-rich star in the stellar stream.
Considering the contribution of s-process, its abundances can be
fitted. The s-process elements in CS 29513-032 are a result of
pollution from the dredged-up material in the former low mass AGB
star (e.g. 1.5M$_\odot$). Base on the analysis of angular momentum
components, \cite{roe10} suggested s-rich star CS 29513-032 is a
member of the stream stars. We find that its r-process
coefficients are close to other stream stars with similar
metallicity, which means that the initial abundance (i.e. before
polluted by the AGB star) of CS 29513-032 is similar to the
abundances of the other stream stars. This implies that the
astrophysical origin of CS 29513-032 and other stream stars is
similar. Our result supports the suggestion by \cite{roe10}. CS
29513-032 should belong to wide binary system which form in a
molecular cloud that have been polluted by both weak and main
r-process material.

Clearly, it is important for future studies to determine the
relation between metal-poor halo stars and metal-poor stellar stream
stars. More theoretical and observational studies of stream stars
will reveal the characteristics of the r-process at low metallicity
and the history of the enrichment of neutron-capture elements in the
early Galaxy.


\begin{table}[h]
\begin{center}
\caption{The component coefficients and $\chi^2$ for 8 stars of the
stellar stream.}\label{table1}
\begin{tabular}{lcccc}
\hline Star & [Fe/H] & $C_{r,m}$ & $C_{r,w}$ & $\chi^2$ \\
\hline BD+10 2495  &   -2.31   &   1.81    &   3.77    &   0.99    \\
BD+29 2356  &   -1.59   &   3.86    &   4.06    &   0.47    \\
BD+30 2611  &   -1.50   &   4.49    &   2.83    &   0.61    \\
CD-36 1052  &   -1.79   &   4.89    &   4.58    &   1.44    \\
HD 119516   &   -2.26   &   3.09    &   3.24    &   1.52    \\
HD 128279   &   -2.51   &   0.86    &   3.86    &   3.01    \\
HD 175305   &   -1.73   &   3.42    &   4.41    &   0.82    \\
HD 237846   &   -3.29   &   0.42    &   3.84    &   2.41    \\

\hline
\end{tabular}
\medskip\\
\end{center}
\end{table}

\section*{Acknowledgments} 

We are grateful to the referee for very valuable comments and
suggestions that improved this paper greatly. This work has been
supported by the National Natural Science Foundation of China
under 10673002, 11273011, 10847119, 10778616, 10973006 and
11003002, the Natural Science Foundation of Hebei Provincial
Education Department under grant 2008127, the Science Foundation
of Hebei Normal University under Grant L2007B07, L2009Z04, the
Natural Science Foundation of Hebei Province under Grant
A2009000251, A2011205102, and the Program for Excellent Innovative
Talents in University of Hebei Province under Grant CPRC034.


\end{document}